\documentclass[12pt,english,round]{article}
\usepackage[T1]{fontenc}
\usepackage[latin9]{inputenc}
\usepackage{geometry}
\usepackage{color}
\usepackage{textcomp}
\usepackage{amsmath}
\usepackage{amssymb}
\usepackage{setspace}
\usepackage[authoryear]{natbib}
\makeatletter

\newcommand*\LyXThinSpace{\,\hspace{0pt}}

\usepackage{url}

\makeatother

\usepackage{babel}
\begin{document}
\title{The Role of Uncertainty in Controlling Climate Change}
\author{\vspace{1cm}
Yongyang Cai\thanks{Department of Agricultural, Environmental and Development Economics,
The Ohio State University, 2120 Fyffe Road, Columbus, OH, USA 43210.
cai.619@osu.edu. }}
\maketitle

\subsection*{Summary and Keywords}

Integrated Assessment Models (IAMs) of the climate and economy aim
to analyze the impact and efficacy of policies that aim to control
climate change, such as carbon taxes and subsidies. A major characteristic
of IAMs is that their geophysical sector determines the mean surface
temperature increase over the preindustrial level, which in turn determines
the damage function. Most of the existing IAMs assume that all of
the future information is known. However, there are significant uncertainties
in the climate and economic system, including parameter uncertainty,
model uncertainty, climate tipping risks, and economic risks. For
example, climate sensitivity, a well-known parameter that measures
how much the equilibrium temperature will change if the atmospheric
carbon concentration doubles, can range from below one to over ten
in the literature. Climate damages are also uncertain: some researchers
assume that climate damages are proportional to instantaneous output,
while others assume that climate damages have a more persistent impact
on economic growth. The spatial distribution of climate damages is
also uncertain. Climate tipping risks represent (nearly) irreversible
climate events that may lead to significant changes in the climate
system, such as the Greenland ice sheet collapse, while the conditions,
probability of tipping, duration, and associated damage are also uncertain.
Technological progress in carbon capture and storage, adaptation,
renewable energy, and energy efficiency are uncertain too. Future
international cooperation and implementation of international agreements
in controlling climate change may vary over time, possibly due to
economic risks, natural disasters, or social conflict. In the face
of these uncertainties, policymakers have to provide a decision that
considers important factors such as risk aversion, inequality aversion,
and sustainability of the economy and ecosystem. Solving this problem
may require richer and more realistic models than standard IAMs, and
advanced computational methods. The recent literature has shown that
these uncertainties can be incorporated into IAMs and may change optimal
climate policies significantly.

Keywords: climate policy, parameter uncertainty, economic risk, climate
risk, model uncertainty, scenario uncertainty, ambiguity, misspecification,
policy uncertainty

\newpage{}

\section*{Uncertainty and Climate Policy}

It has been widely recognized that anthropogenic greenhouse gas emissions
have been distorting the planet's energy balance, resulting in global
warming, sea level rise, and more frequent extreme weather, with industrial
carbon emissions constituting the major component of greenhouse gas
emissions.\footnote{See \citet{hsiang_economists_2018} for an introduction to the physical
science of climate change for economists.} These emissions then influence economic well-being via a damage function.
Integrated Assessment Models (IAMs) combine the climate and the economy,
as well as the interactions between them, to analyze which policies
are more efficient in controlling climate change. DICE \citep{Nordhaus_DICE2007,Nordhaus_DICE2016}
is a representative IAM. Most existing IAMs assume that the climate
and economic systems as well as interactions between them are deterministic,
and that economic agents are myopic. However, there are significant
uncertainties in these systems and their interactions, and these uncertainties
may play an essential role in determining optimal policy. \textcolor{black}{For
example, the dismal theorem of }\citet{Weitzman2009_Dismal}\textcolor{black}{{}
shows that the risk premium could be infinite for unboundedly distributed
uncertainties.}\footnote{\textcolor{black}{See \citet{chari_role_2018} for a discussion about}
uncertainty and risk in climate change.} \citet{pindyck_climate_2013,pindyck_use_2017} criticizes IAMs as
being crucially flawed and fundamentally misleading, but \citet{heal_economics_2017}
argues that ``IAMs can play a role in providing qualitative understanding
of how complex systems behave, but are not accurate enough to provide
quantitative insights. Arguments in favor of action on climate issues
have to be based on aversion to risk and ambiguity and the need to
avoid a small but positive risk of a disastrous outcome.''

However, a policymaker often has to make a quantitative decision,
such as the size of a carbon tax, in the face of this uncertainty.
Moreover, because carbon emissions have long-lasting impacts on temperature,
the ``wait and see'' approach may not make sense given that changes
in the climate system may be irreversible. For example, \citet{steffen_trajectories_2018}
point out a risk that if the Earth System crosses a planetary threshold
then continued warming could occur even as human emissions are reduced,
preventing climate stabilization. Moreover, once an ice sheet collapses,
it is irreversible for millennia \citep{IPCC2013}. \citet{metcalf_integrated_2017}
argue that policymakers need a numerical value for the social cost
of carbon (SCC) for U.S. regulatory policy evaluation and implementation,
and producing a credible numerical value requires sophisticated computer
models, i.e. IAMs.\footnote{\citet{dietz_environmental_2010} suggest that when facing an emission
quantity target (e.g., determined through the Paris Agreement for
keeping a globally average atmospheric temperature increase this century
well below 2°C over the preindustrial level), the marginal abatement
cost (i.e. the shadow price of the target constraint), rather than
the SCC, will often provide more consistent and robust prices for
achieving the target. } \citet{brock_wrestling_2018} stress: ``Defenses for policies that
combat climate damage externalities induced by human activity need
not require precise knowledge of the magnitude or timing of the potential
adverse impacts. ...Waiting for precise knowledge of the eventual
consequences of continued or expanded human induced CO2 emissions
could make mitigation or adaptation extremely costly.'' \citet{goulder_timing_2020}
calls for urgent and stronger policy action to address global climate
change.\footnote{\citet{kolstad_learning_1996} suggests that with the tension between
postponing control until more is known vs acting now before irreversible
climate change takes place, a temporary carbon tax may dominate a
permanent one because a temporary tax may induce increased flexibility
to future uncertainty. }

This review focuses on recent work about the role of uncertainty in
controlling climate change. Here I use a broad perspective of uncertainty,
which includes parameter uncertainty, risk, model uncertainty, scenario
uncertainty, policy uncertainty, ambiguity, and misspecification.
I focus on discussing the first five types of uncertainty and methods
of making decisions in the face of these uncertainties (see \citet{brock_wrestling_2018}
for a more complete survey about ambiguity and misspecification).
Moreover, I focus on reviewing discrete-time stochastic IAMs (see
\citet{brock_chapter_2018} for discussion of continuous-time IAMs).

The rest of the article is organized as follows. It begins with a
discussion of parameter uncertainty and how to deal with it, focusing
on uncertainty in the discount rate, climate sensitivity, and damage
function. This discussion is followed by a review of economic and
climate risks and methods for handling them. The review then goes
on to discuss model uncertainty, scenario uncertainty, ambiguity and
misspecification, as well as policy uncertainty. . Finally conclusion
and issues for further research are presented.

\section*{Parameter Uncertainty\label{sec:Param-uncer}}

Parameters in IAMs are estimated from historical data, expert opinions,
survey data, and/or projections for future scenarios. Their values
are often uncertain because no model can replicate the real world,
historical data may have errors, expert opinions and survey data are
subjective, or projections for future scenarios may be not close to
what will actually happen. These uncertain parameters are assumed
to have fixed and unchanged values over time, but their exact values
are unknown. In some cases, knowledge of the exact values can be expressed
by some probability distributions, which are referred to as belief
distributions. In this review, parameter uncertainty represents only
the cases in which an uncertain parameter has an unknown true value
that is unchanged over time, in order to distinguish with risk representing
the cases with different realization over time and simulation. If
an uncertain parameter's true values have a linear time trend, then
it can be decomposed into two parameters: intercept and slope, both
of which can have parameter uncertainty. Similarly for an uncertain
parameter with a nonlinear time trend true values.\footnote{In fact, usually a time varying path would not be called a parameter. } 

The most well-known and debated uncertain parameters in IAMs include
the discount rate, climate sensitivity (also known as ``equilibrium
climate sensitivity''), and parameters in climate damage functions.
These parameters can change optimal solutions significantly but are
still hard to pin down. Other uncertain parameters include economic
growth, the intertemporal elasticity of substitution, and risk aversion.\footnote{See e.g. \citet{Gillingham_etal_2018,CaiJuddLontzek2018_Comp,CaiLontzek2019_DSICE}
for investigations on the impact of these uncertain parameters.}

\subsection*{\textit{Discount Rate}}

Economic analysis often uses two types of discount rates: a utility
discount rate\footnote{It has other names like the pure rate of time preference.}
that represents the rate at which utility is discounted, and a consumption
discount rate\footnote{Other names include the social rate of time preference and the social
discount rate.} that represents the rate at which consumption is discounted. These
two discount rates can be connected by the famous Ramsey rule. There
are a large number of papers discussing these discount rates, see
for example \citet{weitzman_gamma_2001}, \citet{frederick_time_2002},
\citet{Gollier2012}, \citet{arrow_determining_2013,arrow_etal_2014},
\citet{heal_economics_2017}, and \citet{drupp_discounting_2018}.

\citet{IWG2010} employs three consumption discount rates (2.5\%,
3\%, and 5\%) to compute the SCC. \citet{Nordhaus_DICE2007} uses
a utility discount rate of 1.5\%, which is calibrated together with
the intertemporal elasticity of substitution to match the estimated
growth of consumption. For ethical reasons, \citet{Stern2007} sets
a utility discount rate of 0.1\% and finds that the SCC will be significantly
higher.

\subsection*{\textit{Climate Sensitivity and Transient Climate Response}}

Climate sensitivity, also known as equilibrium climate sensitivity
(ECS), is a parameter that measures the long-run increase in atmospheric
temperature (in degrees Celsius) if the atmospheric carbon concentration
doubles. A typical value of climate sensitivity used in the literature
is 3°C, which is considered to be the median of the distribution of
climate sensitivity. \citet{IPCC2007} suggests the likely range (i.e.
with a 66\% probability) of climate sensitivity is $[2.0,4.5]$, but
a later report \citet{IPCC2013} expands the likely range to $[1.5,4.5]$
(the same as given by Jule Charney in 1979) in light of recent research,
demonstrating that it is challenging to narrow the envelope of parameter
uncertainty of climate sensitivity with additional research. \citet{meinshausen_greenhouse-gas_2009}
plot 19 probability density distributions of climate sensitivity,
representing the wide variety of climate modeling approaches, observational
data, and statistical methodologies in the literature. The range of
the ECS value is from below one to over ten, and some distributions
are skew to the left while others are skew to the right. 

Instead of climate sensitivity, recently some climate scientists suggest
using a more stable measurement, called ``transient climate response
to emissions'' (TCRE), to simplify modeling the climate system. The
TCRE scheme assumes that contemporaneous atmospheric temperature increase
is nearly linearly proportional to cumulative carbon emissions. Therefore,
atmospheric temperature can be obtained with only one state variable,
cumulative emissions. In contrast, the climate system modeled using
climate sensitivity is often complicated and requires many state variables.\footnote{For example, DICE uses five state variables for the climate system:
carbon concentration in the atmosphere, carbon concentration in the
upper ocean, carbon concentration in the deep ocean, atmospheric temperature,
and oceanic temperature. See \citet{matthews_proportionality_2009},
\citet{macdougall_origin_2015}, and \citet{knutti_beyond_2017} for
further details.} With the simplification of the TCRE scheme, economists have also
employed it in their models, including \citet{brock_climate_2017},
\citet{Anderson_Robust_2018}, and \citet{van_der_ploeg_safe_2018}.
Recently, \citet{dietz_cumulative_2019} use a continuous-time IAM
with the TCRE scheme and find that the optimal carbon price should
start relatively high and grow relatively fast. However, the value
of TCRE is still uncertain and highly correlated to ECS. \citet{macdougall_uncertainty_2016}
estimate the mean value of TCRE as 1.72°C (that is, if there are additional
100 gigaton of carbon (GtC) emissions, then the mean atmospheric temperature
increases 1.72 °C), and its 5\% to 95\% percentile range as {[}0.88,
2.52{]}, while \citet{IPCC2013} reports its likely range as {[}0.8,
2.5{]}.

\subsection*{\textit{Damage Function}}

The specification of the damage function, which measures the damage
from global warming, has been debated extensively. The most common
damage function is a quadratic function of the temperature increase,
specified by Professor William Nordhaus in his DICE/RICE models \citep{Nordhaus_DICE2007,Nordhaus_DICE2016,Nordhaus_RICE_2010}.
It assumes that temperature increases reduce instantaneous economic
output in a ratio represented by the damage function. However, \citet{Weitzman2012}
points out that the quadratic function results in implausibly low
damage at high temperatures, and thus suggests adding one high-exponent
term to the damage function so that 50\% of output is lost if the
temperature increase is 6 °C. \citet{DietzStern2015} find that the
modification of \citet{Weitzman2012} leads to a much higher SCC.
\citet{diaz_quantifying_2017} review and synthesize the limitations
of the damage functions.

\citet{sterner_even_2008} argue that ``it is exactly the nonmarket
effects of climate change that are the most worrisome'', which include
``biodiversity and ecosystem loss; effects on human well-being (human
amenity, loss of lives, and air pollution); impacts from natural disasters,
such as extreme weather events, droughts, hurricanes or floods \citep{manne_merge_1995};
as well as socially contingent consequences, such as migration and
risk for conflicts'', many of which have been presented in the Stern
review \citep{Stern2007}. The nonmarket amenities and aggregate consumption
can be combined with a utility function with a constant elasticity
of substitution kernel, which leads to a much higher SCC \citep{sterner_even_2008,Cai_PNAS2015}.
That is, explicitly including the nonmarket amenities in the utility
function allows for relative price changes and then leads to very
different results, as compared with implicitly including the nonmarket
amenities as the equivalent loss in consumption of market goods.

Besides the climate damage to instantaneous output and nonmarket goods,
researchers also find that climate change can reduce economic growth.
For example, \citet{Dell2012} find a reduction in economic growth
of approximately 1.3\% for a 1°C increase in global temperatures.
In the face of the damage to economic growth, the SCC will be significantly
higher \citep{moore_NCC_2015,DietzStern2015}.\footnote{See \citet{heal_reflectionstemperature_2016} for a review about the
climate impact on output levels and growth rates.} Some recent empirical works in estimating climate impacts include
\citet{burke_climate_2015,burke_nature_2015}, \citet{burke_incorporating_2015},
\citet{carleton_social_2016}, \citet{costinot_evolving_2016}, \citet{hsiang_estimating_2017},
\citet{schauberger_consistent_2017}, \citet{burke_large_2018}, \citet{chen_coastal_2018},
\citet{fan_climate_2018}, \citet{schlenker_cost_2018}, \citet{zhang_temperature_2018},
\citet{diffenbaugh_global_2019}, \citet{duffy_strengthened_2019},
\citet{hsiang_distribution_2019}, and \citet{mach_climate_2019}.
For example, \citet{burke_nature_2015} find that overall economic
productivity is nonlinear in temperature, and then estimate that in
2100 ``unmitigated climate change will make 77\% of countries poorer
in per capita terms than they would be without climate change'',
and their estimate of damage in global GDP is much larger than those
from major IAMs including DICE \citep{Nordhaus_DICE2007}. 

\subsection*{\textit{Dealing with Parameter Uncertainty}}

The most common way to deal with parameter uncertainty is sensitivity
analysis, that is, choosing a different value of the uncertain parameter
and checking its impact on the solution to see if results are qualitatively
robust. If there are multiple uncertain parameters, then doing sensitivity
analysis over each uncertain parameter might not be enough, as a combination
of multiple uncertain parameters may produce nontrivial results. Thus,
a global sensitivity analysis\textemdash going through a number of
combinations of multiple uncertain parameters\textemdash should be
conducted. For example, \citet{CaiJuddLontzek2017_DSICE} and \citet{CaiLontzek2019_DSICE}
use global sensitivity analysis in a dynamic stochastic integrated
framework of climate and economy (DSICE) to study the impact of the
intertemporal elasticity of substitution (IES) and risk aversion on
the SCC in the face of long-run economic growth risk. They find that
a larger degree of risk aversion increases the SCC if the IES is small,
but decreases the SCC if the IES is large, which can be reconciled
using the impact of economic growth. However, in the face of climate
tipping risk, a larger degree of risk aversion always increases the
SCC regardless of the size of the IES. It may be too time-consuming
to run global sensitivity analysis if tensor grids for uncertain parameters
are used. Uncertainty quantification can be applied to address this
issue: choose a small set of nodes (e.g. a sparse grid) for uncertain
parameters, then apply an approximation method to estimate solutions
over the whole domain of uncertain parameters (see \citet{harenberg_uncertainty_2019}
for a detailed discussion).

Sensitivity analysis methods provide the lower and upper bounds of
the solutions and also the qualitative robustness of solutions, but
it is often challenging to provide one specific quantitative value,
e.g. the SCC, for use in decision making, as policymakers do not know
which values of uncertain parameters are correct. However, if policymakers
have a belief distribution for possible values of the uncertain parameters,
then it is possible to give a quantitatively robust solution. For
example, \citet{CaiSanstad2016} use an expected cost minimization
method to find a robust mitigation pathway in the face of R\&D (research
and development) technology uncertainty. Some researchers use a Monte
Carlo method: they draw samples of the uncertain parameters from the
belief distributions, solve the deterministic model with each sampled
realization of the uncertain parameters, then use the average over
the solutions as an approximate solution in the face of uncertainty.\footnote{See e.g., \citet{NewHulme2000}, \citet{Nordhaus_DICE2007}, \citet{Ackerman_etal2010},
and \citet{AnthoffTol2013}.} While this Monte Carlo analysis may be helpful in some cases, it
may also lead to very biased solutions (see \citet{CrostTraeger2014},
\citet{LemoineRudik2017} and \citet{Cai2019} for detailed discussions).

If policymakers have belief distributions of uncertain parameters,
they may collect more data over time to update their belief distributions,
by shrinking the range of uncertain parameters or reducing the variances
of these distributions. This process is called Bayesian learning.
Note that any sampled realization of the uncertain parameters in simulated
solutions should be their fixed true values, so the true values have
to be assumed before simulation in order to study the impact of learning
under the assumed true values. Bayesian learning has been applied
in climate change economics, for example in \citet{KellyKolstad1999},
\citet{keller_uncertain_2004}, \citet{Leach2007}, \citet{KellyTan2015},
\citet{hwang_effect_2017}, \citet{gerlagh_carbon_2018}, and \citet{rudik_optimal_2019}.
Since Bayesian learning depends on the frequency of new data collection,
in dynamic models it depends on the size of the time steps: with annual
time steps belief distributions can be updated every year, but with
decadal time steps belief distributions can be updated only every
decade. Thus, the size of the time steps should be chosen reasonably
to be consistent with the frequency of real data collection.\footnote{In dynamic programming models, the frequency of policy updates in
the real world should also be taken into account the choice of the
time step size. For example, climate policies might not update annually,
while consumption decisions could be more frequent, so an IAM with
annual time steps may be more suitable than those with decadal time
steps or continuous time. See \citet{cai_open_2012} for an example
that a solution with decadal time steps is significantly different
with one with annual time steps. } In addition, Bayesian learning ignores the possibility of active
learning through R\&D (see \citealt{GoulderMathai2000}), so it may
provide biased results.

In the face of parameter uncertainty without belief distributions,
decision makers may have to provide only one robust solution instead
of many solutions from (global) sensitivity analysis. This robust
decision making problem can be solved using a robust decision making
method, such as the max-min method or the min-max regret method \citep{Iverson2012,Iverson2013,AnthoffTol2014,CaiSanstad2016}.
See \citet{Cai2019} for a detailed discussion.

\section*{Risks\label{sec:risk}}

There are many risks in economic and climate systems, where risks
refer to random variables with probability distributions that are
known or dependent on state or control variables at each time period
but could be time-varying. Unlike parameter uncertainty where the
true value of an uncertain parameter is unchanged over time, risks
are assumed to be realized with possibly different values over time
and simulation, and a different simulation could lead to a different
time varying path. Risks in the economic system can happen in technology,
productivity, health and mortality, research and development (R\&D),
and international cooperation and noncooperation, as well as in many
other areas. Risks in the climate system include the frequency and
damage of extreme weather, and regime switching of the climate system
(often called climate tipping risks). In particular, fat-tailed uncertainty
in catastrophic climate change is the most worrisome \citep{weitzman_fat-tailed_2011}.

\subsection*{\textit{Economic Risks}}

The economic system of an IAM often assumes that output is proportional
to total factor productivity (TFP), $A_{t}$, and business cycle models
assume that $A_{t}$ is a stochastic process. For example, \citet{fischer_emissions_2011}
use a dynamic stochastic general equilibrium (DSGE) model with $\ln(A_{t+1})=\rho\ln(A_{t})+\epsilon_{t}$,
where $\epsilon_{t}$ is an i.i.d. shock which is normally distributed
with mean zero, to compare the performance of three instruments in
achieving an exogenous and fixed level of expected emissions reduction:
a carbon tax, emissions cap-and-trade (i.e. an exogenous limit on
aggregate emissions), and an emission intensity target (i.e. an exogenous
limit on emissions per unit of aggregate output). They find that the
intensity target scheme produces higher mean values and lower welfare
costs than the other two policies, and compared to the no-policy case,
volatility of the main macroeconomic variables decreases under the
carbon tax policy but increases under the cap-and-trade instrument.
\citet{heutel_2012} also uses the same form of $A_{t}$ in his DSGE
model, but focuses on comparing the optimal emissions tax rate and
the optimal emissions quota. He finds that they both decrease during
recessions, and during economic expansions a price effect from costlier
abatement dominates an income effect of greater demand for clean air.
\citet{fischer_environmental_2013} survey some related work using
real business cycle models with environmental policy and induced technological
progress.

\citet{annicchiarico_2015} formulate a New-Keynesian-type DSGE model
to study the role of different environmental policy regimes in economic
fluctuations, in the presence of nominal rigidities and accounting
for two additional sources of uncertainty: public consumption shocks
and monetary policy shocks. They find that in the presence of nominal
rigidities, a cap-and-trade scheme is likely to dampen the response
of the main macroeconomic variables to shocks, an intensity target
makes macroeconomic variables more volatile, and a carbon tax policy
tends to have slightly higher mean welfare and lower volatility than
the cap-and-trade scheme, as long as the degree of price stickiness
is not too high. However if prices adjust very slowly, the cap-and-trade
policy will have higher mean welfare. \citet{karydas_pricing_2019}
include transition risks of climate policy (i.e. the risks associated
with carbon-intensive assets, which may become stranded due to stringent
climate policies) in their dynamic asset pricing framework with rare
disasters related to climate change, and show that transition risks
substantially lower the participation of carbon-intensive assets in
the market portfolio.

\citet{JensenTraeger2014}, \citet{CaiJuddLontzek2017_DSICE}, and
\citet{CaiLontzek2019_DSICE} also build DSGE models assuming that
TFP, $A_{t}$, is a Markov process but is extended to be a long-run
risk \citep{BansalYaron2004} that includes persistence of shocks.
That is, they assume $A_{t}=\widetilde{A}_{t}\zeta_{t}$ where $\widetilde{A}_{t}$
is an exogenous deterministic trend of TFP from DICE, and $\zeta_{t}$
is a shock with \textcolor{black}{
\begin{equation}
\log\left(\zeta_{t+1}\right)=\log\left(\zeta_{t}\right)+\chi_{t}+\varrho\omega_{\zeta,t}\label{eq:zeta_process}
\end{equation}
\begin{equation}
\chi_{t+1}=r\chi_{t}+\varsigma\omega_{\chi,t},\label{eq:chi_process}
\end{equation}
where $\chi_{t}$ represents the persistence of the shock, $\omega_{\zeta,t},\omega_{\chi,t}\sim i.i.d.\:\mathcal{N}(0,1)$,
and $\varrho$, $r$, and $\varsigma$ are parameters.} Moreover,
the papers incorporate a recursive utility function \citep{Epstein-Zin-1989}
with two preference parameters: the intertemporal elasticity of substitution
(IES), and relative risk aversion. However, the DSICE model of \citet{CaiJuddLontzek2017_DSICE}
and \citet{CaiLontzek2019_DSICE} is a stochastic extension of the
full DICE model \citep{Nordhaus_DICE2007}, and its parameters for
the long-run risk on growth are calibrated with historical consumption
growth data, whereas \citet{JensenTraeger2014} use a reduced form
model without \textcolor{black}{empirical validation} for the calibration
of their long-run risk (see \citet{CaiJuddLontzek2017_DSICE} and
\citet{CaiLontzek2019_DSICE} for a more detailed discussion). Moreover,
\citet{CaiJuddLontzek2017_DSICE} and \citet{CaiLontzek2019_DSICE}
discretize $\zeta_{t}$ and \textcolor{black}{$\chi_{t}$ to be Markov
chains with a large number of time-varying values, while }\citet{JensenTraeger2014}\textcolor{black}{{}
do not. The discretization is to avoid existence issues caused by
the unbounded normal distributions of $\omega_{\zeta,t}$ and $\omega_{\chi,t}$,
and to avoid excessive dependence on extreme tail events (as mentioned
in Section 1, the risk premium could be infinite for unboundedly distributed
uncertainties} \citep{Weitzman2009_Dismal}\textcolor{black}{).} \citet{CaiJuddLontzek2017_DSICE}
and \citet{CaiLontzek2019_DSICE} find that in the presence of long-run
risk on growth, the SCC itself is a stochastic process with a wide
range of possible values, and the recursive utility's preference parameters
have a nontrivial impact on the SCC in 2005: if the IES is large (not
less than 0.9 in their numerical examples), then a larger risk aversion
implies a smaller SCC; if IES is small (not larger than 0.7 in their
numerical examples), then a larger risk aversion implies a larger
SCC; and if risk aversion is small (not larger than 5), then the SCC
increases with the IES.\footnote{Asset pricing theory has also been applied to estimate the SCC in
the face of risks. For example, \citet{bansal_climate_2018} and \citet{daniel_applying_2018}
explore the implications of risk preferences for the SCC and optimal
abatement policies. \citet{bansal_climate_2019} show that the long-run
temperature elasticity of equity valuations is significantly negative
and that long-run temperature fluctuations carry a positive risk premium
in equity markets.}

\subsection*{\textit{Climate Risks}}

\textcolor{black}{One major type of climate risk is `climate tipping
risks', referring to risks that can qualitatively alter the state
or development of the climate system, if a large-scale component of
the Earth system (the `tipping element') passes a critical threshold
(the `tipping point'). Some examples of tipping elements include the
Greenland ice sheet, the Antarctic ice sheet, and the Amazon rainforest.
A climate tipping risk is often represented by an irreversible climate
process, called a }\textit{tipping process}\textcolor{black}{, which
is often a Markov chain.}\footnote{\textcolor{black}{See e.g. \citet{lenton_tipping_2008}, \citet{kriegler_imprecise_2009},
\citet{scheffer_early-warning_2009}, \citet{ditlevsen_tipping_2010},
and \citet{ghil_physics_2019} for a discussion about the physics
and early warning of climate tipping points.}}

\citet{LemoineTraeger2014} study the impact of tipping points on
climate policy, where their tipping point is an instantaneous, irreversible
increase in climate sensitivity (from 3°C to 4, 5, or 6°C), or an
instantaneous, irreversible weakening of carbon sinks (by 25, 50,
or 75\%). However, \citet{lontzek_NCC_2015} point out that such a
\textquoteleft tipping point\textquoteright{} is not scientifically
plausible, because \textcolor{black}{``positive feedbacks are never
instantaneously switched on \textendash{} instead they may get progressively
stronger as temperature increases \textendash{} so an instantaneous
\textquoteleft tipping\textquoteright{} formulation is qualitatively
wrong.'' Climate scientists point out that the duration of a climate
element tipping to a new state is unknown and can last from a decade
to even millennia \citep{lenton_integrating_2013}. There is no widespread
belief in the scientific community that the entire Greenland ice sheet
will melt within one year and cause an instant global sea level rise
of six meters.}

\citet{LemoineTraeger2014} \textcolor{black}{assume that a tipping
point cannot occur when the contemporaneous temperature is below last
period's temperature.}\footnote{\textcolor{black}{Footnote 5 of }\citet{LemoineTraeger2014}\textcolor{black}{{}
states: ``}\textit{\textcolor{black}{In our climate application,
the decision maker keeps track of the greatest historic temperature}}\textcolor{black}{.''
However, given their model equations and code, that statement is wrong.
The decision maker keeps track of the }\textit{\textcolor{black}{last-period}}\textcolor{black}{{}
temperature.}}\textcolor{black}{{} However, this assumption violates the consensus
in climate science and completely ignores recent findings that suggest
climate tipping might already have occurred, despite the last two
decades, in which global warming had a short decreasing trend. For
example, there is already scientific evidence that major ice sheets
are losing mass at an accelerating rate (\citealt{khan_sustained_2014}).
The Greenland ice sheet's mass loss is estimated to be contributing
\textasciitilde 0.7 mm/year to sea level rise \citep{csatho_laser_2014},
and \citet{joughin_marine_2014} argue that the collapse of the West
Antarctic ice sheet is already underway. Moreover, a real temperature
path always rises and falls frequently, so the tipping probabilities
assumed by \citeauthor{LemoineTraeger2014} will also frequently fluctuate
between nonzero and zero, which is implausible. Furthermore, }\citet{LemoineTraeger2014}\textcolor{black}{{}
assume that the tipping probability only depends on the positive change
between this year's and next year's temperature. Hence, no matter
how low the current year's temperature is, as long as it increases,
the tipping probability will be nonzero. Moreover, no matter how high
the current year's temperature is, if next year's temperature does
not increase, then the tipping probability is zero. Thus, the decision-maker's
mitigation efforts and optimal climate policy will be based on distorted
incentives. }

\textcolor{black}{In addition, the hazard rate in }\citet{LemoineTraeger2014}\textcolor{black}{{}
is not based on a calibration in part because climate scientists do
not know when or at which level of global warming a tipping point
will occur \citep{lenton_tipping_2008,kriegler_imprecise_2009}. }\citet{LemoineTraeger2014},
however,\textcolor{black}{{} assume that a tipping point will occur
if temperatures reach 4.27°C above preindustrial levels. When one
aggregates the implied probabilities from expert elicitation by \citet{kriegler_imprecise_2009},
one finds that tipping is more likely than not in a 4\textendash 8°C
long-term warming scenario, but still not certain.}

\textcolor{black}{In contrast, }\citet{CaiJuddLontzek2017_DSICE}
and \textcolor{black}{\citet{CaiLontzek2019_DSICE} use an alternative
approach to investigate the impact on the SCC from climate tipping
risks. Their tipping point process incorporates basic features of
how climate scientists think about climate tipping points, such as
a stochastic formulation of the physical process of triggering the
tipping point and a transition time of tipping impacts with uncertain
duration and long-run impact size. }Their hazard rate formulation
treats the physical process of the tipping point as stochastic with
tipping probabilities that depend on the contemporaneous temperature
itself, so a higher temperature will always have a higher tipping
probability. \textcolor{black}{Their hazard rate is calibrated according
to beliefs expressed in expert elicitation studies, where the experts
based their probability statements partly on the Earth's history,
partly on fundamental understanding, and partly on future model projections
\citep{lenton_tipping_2008}. }\citet{CaiJuddLontzek2017_DSICE} and
\textcolor{black}{\citet{CaiLontzek2019_DSICE} model their tipping
process as a 5-stage sequential process with each stage having a stochastic
duration. Thus, not only do they model the long transition of climate
tipping, as postulated by climate scientists, but they also account
for these scientists' lack of knowledge and imperfect information
regarding the length of the transition process. Their multiple-stage
approach to modeling the representative tipping element can produce
lengths of transition in accordance with scientific beliefs. }\citet{CaiJuddLontzek2017_DSICE}
and \textcolor{black}{\citet{CaiLontzek2019_DSICE} also account for
the additional uncertainty about the long-run impact of climate tipping
on the economy, which according to climate scientists, is the biggest
uncertainty.}

\citet{CaiJuddLontzek2017_DSICE} and \textcolor{black}{\citet{CaiLontzek2019_DSICE}
find that under }recursive utility and \textcolor{black}{climate tipping
risks, a higher IES or risk aversion always implies a higher SCC in
2005. If a tipping event has not happened, then the SCC is significantly
higher than in a deterministic model, but it will jump down significantly
and immediately once the tipping event happens, even though the post-tipping
damage has just started to unfold and may take many years to reach
its maximum level. This pattern appears because the incentive to }prevent
or delay the tipping event has disappeared if it has been triggered.

The DSICE framework has also been applied with various variants in
the literature. \citet{lontzek_NCC_2015} use it to investigate the
impact of a tipping risk with a continuous tipping damage path under
separable utility, and find that the  optimal carbon tax in 2005 increases
by around 50\% even with conservative assumptions about the rate and
impacts of a stochastic tipping event. Moreover, the effective discount
rate for the costs of stochastic climate tipping is much lower than
the discount rate for deterministic climate damages. \citet{Cai_PNAS2015}
use the DSICE framework to study the impact of environmental tipping
risk on market and nonmarket goods and services, and find that even
if a tipping risk only has nonmarket impacts, it could substantially
increase the optimal carbon tax in 2005. \citet{cai_NCC_2016} extend
DSICE to incorporate five major interacting climate tipping risks
(Atlantic meridional overturning circulation, disintegration of the
Greenland ice sheet, collapse of the West Antarctic ice sheet, dieback
of the Amazon rainforest, and shift to a more persistent El Niño regime)
simultaneously in their model. They find that doing so increases the
SCC in 2005 by nearly eightfold, and passing a tipping point may abruptly
increase the SCC if it increases the likelihood of other tipping events
(note that if there is only one tipping risk, then passing its tipping
point will abruptly decrease the SCC, as shown in \citet{lontzek_NCC_2015},
\citet{Cai_PNAS2015}\textcolor{black}{, and \citet{CaiLontzek2019_DSICE}}). 

\citet{diaz_potential_2016} find that the threat of the West Antarctic
ice sheet disintegration implies little motivation for additional
mitigation. \citet{nordhaus_economics_2019} finds that the risk of
Greenland ice sheet disintegration makes a small contribution to the
overall social cost of climate change. These findings are consistent
with \citet{cai_NCC_2016} in the case without interactions between
tipping events (see Figure 3 of \citet{cai_NCC_2016}). But \citet{cai_NCC_2016}
also show that in the case with interactions between tipping events,
if the Greenland ice sheet tips first, it leads to the most stringent
emissions control because the likelihood of the presumably most-damaging
event (Atlantic meridional overturning circulation collapse) significantly
increases. \citet{anthoff_shutting_2016} apply the FUND (Climate
Framework for Uncertainty, Negotiation and Distribution) model to
study the impact of potential slowdown of the thermohaline circulation
(i.e., Atlantic meridional overturning circulation), a vast system
of currents across all four oceans. They find that the change in human
welfare associated with a slowdown of the thermohaline circulation
is modest, if only its potential to reduce both ocean heat and carbon
uptake is considered but its other effects, such as ocean acidification
and possibly more extreme weathers, are ignored.\footnote{\citet{nordhaus_expert_1994} and \citet{nordhaus_warming_2000} suggest
that a collapse of the thermohaline circulation might result in a
25-30\% reduction in GDP.} 

\citet{Cai_etal2019_DIRESCU} construct a numerical dynamic stochastic
two-region model, DIRESCU, which separates the world into the North
and the Tropic-South regions with their own economic and climate systems
and includes interactions between regions. They also consider a global
climate tipping risk, global sea level rise, and regional adaptation,
and find that carbon taxes increase significantly in both regions
to curb or delay the occurrence of climate tipping and sea level rise
in both cooperative and non-cooperative worlds, while regional adaptation
reduces carbon taxes significantly.

There is also theoretical analysis for pricing carbon in the face
of climate tipping. For example, \citet{van_der_ploeg_non-cooperative_2016}
build a simple stylized North-South model of the global economy to
investigate how differences between regions in terms of their vulnerability
to climate change and their stage of development affect cooperative
and non-cooperative responses, both to curb the risk of a calamity
and accumulate precautionary capital to facilitate consumption smoothing.
\citet{van_der_ploeg_climate_2018} find that if the mean lag for
the impact of the catastrophe from climate tipping is long enough,
the saving response will be negative, because the precautionary return
in the Keynes\textendash Ramsey rule becomes negative.

\subsection*{\textit{Dealing with Risks}}

A typical dynamic stochastic IAM can be written as 
\begin{eqnarray}
\max_{\mathbf{a}_{t}\in\mathcal{D}_{t}(\mathbf{x}_{t})} &  & \mathbb{E}\left\{ \sum_{t=0}^{T-1}\beta^{t}u_{t}(\mathbf{x}_{t},\mathbf{a}_{t})+\beta^{T}V_{T}(\mathbf{x}_{T})\right\} \label{eq:sto-DP}\\
\mathrm{s.t.} &  & \mathbf{x}_{t+1}=\mathbf{f}_{t}(\mathbf{x}_{t},\mathbf{a}_{t},\epsilon_{t}),\;t=0,1,...,T-1\nonumber \\
 &  & \mathbf{x}_{0}\;\mathrm{given}\nonumber 
\end{eqnarray}
where $\mathbb{E}$ is the expectation operator over all random variables
$\epsilon_{t}$ for all time $t$, $\mathbf{x}_{t}$ is a vector of
state variables (such as capital, oil stock, carbon concentration
in the atmosphere, and global average atmospheric temperature), $\mathbf{a}_{t}$
is a vector of control variables (also called action or decision variables,
such as consumption and emission mitigation rates), $\beta$ is the
discount factor, $T$ is the terminal time (could be infinite), $u_{t}$
is a utility function, $V_{T}$ is the terminal value function (when
$T=\infty$, this term disappears), $\mathbf{f}_{t}$ is a vector
of functions representing transition laws of state variables, and
$\mathcal{D}_{t}(\mathbf{x}_{t})$ is a feasible set of the control
variables and depends on the state variables at each time $t$. When
the transition law of the $j$-th state variable is deterministic,
$x_{t+1,j}=g_{t,j}(\mathbf{x}_{t},\mathbf{a}_{t})$, it is still denoted
as $x_{t+1,j}=f_{t,j}(\mathbf{x}_{t},\mathbf{a}_{t},\epsilon_{t})=g_{t,j}(\mathbf{x}_{t},\mathbf{a}_{t})+0\cdot\epsilon_{t}$
for convenience. The model (\ref{eq:sto-DP}) can also be transformed
to the following Bellman equation \citep{Bellman1957}:
\begin{eqnarray}
V_{t}(\mathbf{x}_{t})=\max_{\mathbf{a}_{t}\in\mathcal{D}_{t}(\mathbf{x}_{t})} &  & u_{t}(\mathbf{x}_{t},\mathbf{a}_{t})+\beta\mathbb{E}_{t}\left\{ V_{t+1}(\mathbf{x}_{t+1})\right\} \label{eq:Bellman}\\
\mathrm{s.t.} &  & \mathbf{x}_{t+1}=\mathbf{f}_{t}(\mathbf{x}_{t},\mathbf{a}_{t},\epsilon_{t})\nonumber 
\end{eqnarray}
for $t=0,1,...,T-1$, where $\mathbb{E}_{t}$ is the expectation operator
over $\epsilon_{t}$ conditional on time-$t$ information $(\mathbf{x}_{t},\mathbf{a}_{t})$,
and $V_{t}$ is the value function at time $t$.

In economics, consumption $c_{t}$ is often a decision variable (one
element of $\mathbf{a}_{t}$), and a typical utility function is a
power function: 
\[
u_{t}(\mathbf{x}_{t},\mathbf{a}_{t})=\frac{c_{t}^{1-\gamma}}{1-\gamma}
\]
for $\gamma>0$ and $\gamma\neq1$, where $\gamma=1$ is the special
case of logarithmic utility: 
\[
\lim_{\gamma\rightarrow1}\frac{c_{t}^{1-\gamma}-1}{1-\gamma}=\ln(c_{t}).
\]
For a deterministic dynamic model, $\gamma$ is the inverse of intertemporal
elasticity of substitution (IES). For a stochastic model, $\gamma$
is also the relative risk aversion parameter. To disentangle the IES
from risk aversion, recursive utility \citep{Epstein-Zin-1989} has
been employed in IAMs recently, for example, in \citet{JensenTraeger2014},
\citet{cai_NCC_2016}, \citet{CaiLontzek2019_DSICE}, and \citet{Cai_etal2019_DIRESCU}.
The corresponding Bellman equation is 
\begin{eqnarray}
V_{t}(\mathbf{x}_{t})=\max_{\mathbf{a}_{t}\in\mathcal{D}_{t}(\mathbf{x}_{t})} &  & u_{t}(\mathbf{x}_{t},\mathbf{a}_{t})+\beta\mathcal{G}_{t}\left\{ V_{t+1}(\mathbf{x}_{t+1})\right\} \label{eq:Bellman-EZ}\\
\mathrm{s.t.} &  & \mathbf{x}_{t+1}=\mathbf{f}_{t}(\mathbf{x}_{t},\mathbf{a}_{t},\epsilon_{t})\nonumber 
\end{eqnarray}
where 
\[
\mathcal{G}_{t}\left\{ V_{t+1}(\mathbf{x}_{t+1})\right\} \equiv\frac{1}{1-1/\psi}\left(\mathbb{E}_{t}\left\{ \left((1-1/\psi)V_{t+1}(\mathbf{x}_{t+1})\right)^{\frac{1-\gamma}{1-1/\psi}}\right\} \right)^{\frac{1-1/\psi}{1-\gamma}}
\]
with $\psi$ and $\gamma$ as the IES and the risk aversion parameter
respectively.

The most common method to solve the (time-varying) Bellman equation
is value function iteration (VFI). When some state variables are continuous,
value functions $V_{t}$ have to be approximated. An efficient approximation
method is to use complete Chebyshev polynomials (over multi-dimensional
continuous state variables) and associated Chebyshev nodes. It is
essential for approximation errors to be small, and the errors of
the solution should always be checked, otherwise the numerical solution
could be far away from the true solution.

For example, assume that VFI is employed to solve an infinite-horizon
stationary problem, where the true value function $V$ satisfies the
Bellman equation $V=\Gamma(V)$, where $\Gamma$ is the Bellman operator.
Starting with an initial guess $V_{0}$, VFI solves $V_{t}=\Gamma(V_{t-1})$
for $t=1,2,...$ until it converges under a stopping criterion: $\left\Vert V_{t}-V_{t-1}\right\Vert <\varepsilon$,
where $\varepsilon$ is a small positive number and $\left\Vert \cdot\right\Vert $
is a norm operator over functions. Since $V_{t}$ cannot be solved
at all states if some state variables are continuous, numerically
$V_{t}$ is solved at approximation nodes $\mathbf{x}_{j}$ and then
values $V_{t}(\mathbf{x}_{j})$ are used to construct an approximation
of $V_{t}$ at all states. Let the value function at iteration $t$
be approximated by a linear combination of basis functions, denoted
$\widehat{V}_{t}$, and numerical VFI converges under a stopping criterion:
$\left\Vert \widehat{V}_{t}-\widehat{V}_{t-1}\right\Vert <\varepsilon$.
This convergence does not guarantee that $\widehat{V}_{t}$ is a good
approximation to the true value function $V$, as numerical VFI computes
$(\Gamma(\widehat{V}_{t-1}))(\mathbf{x}_{j})$ and then uses the values
$(\Gamma(\widehat{V}_{t-1}))(\mathbf{x}_{j})$ to approximate $\Gamma(\widehat{V}_{t-1})$
with $\widehat{V}_{t}$; that is $\widehat{V}_{t}\neq\Gamma(\widehat{V}_{t-1})$.
In fact, numerical VFI may converge under the stopping criterion with
any degree of approximation (such as linear or quadratic polynomial
approximation). Therefore, if the approximation error, $\left\Vert \widehat{V}_{t}-\Gamma(\widehat{V}_{t-1})\right\Vert $,
is large, then the converged solution $\widehat{V}_{t}$ may be far
away from the true solution. In addition, even if a high-degree Chebyshev
polynomial is used in approximation, a loose stopping criterion may
also lead to large errors in the solution. For example, \citet{Cai2019}
points out that the stopping criterion of VFI used in \citet{LemoineTraeger2014}
is problematic, so the numerical solution of \citet{LemoineTraeger2014}
may have large errors. \footnote{See \citet{Judd1998} and \citet{Cai2019} for details about computational
methods and error checking.}

However, the Bellman equation (\ref{eq:Bellman}) or (\ref{eq:Bellman-EZ})
only applies to a social planner's problem, where the social planner
makes all decisions including reallocating resources among different
regions or countries (with some costs), while these regions/countries
are completely cooperative. In the real world, the regions/countries
may be non-cooperative, which leads to a (time-varying) dynamic stochastic
game. \citet{Cai_etal2019_DIRESCU} introduce a new time-backward
iterative method to solve a system of Bellman equations and then find
a feedback Nash equilibrium numerically.

In many cases, it could be very challenging to employ VFI if there
are kinks in value functions and/or the number of state variables
is large. \citet{Cai_etal_QE2017} introduce a nonlinear certainty
equivalent approximation method to solve an infinite-horizon stationary
problem in the form (\ref{eq:sto-DP}). \citet{Cai_etal2020_ENLCEQ}
then extend it to solve a (finite/infinite) nonstationary problem.
See \citet{Cai2019} for discussion about other computational methods.

\section*{Model Uncertainty and Scenario Uncertainty}

IAMs and scenarios are developed to analyze climate policies and estimate
future pathways of temperature, but no model can replicate the real
world completely, and no scenario can predict a realized future pathway
perfectly. For tractability, every model or scenario has to make some
simplifying assumptions, particularly in mathematical representations
of economic and climate systems. Different assumptions then lead to
different models or scenarios. In the words of Albert Einstein, ``Everything
should be made as simple as possible, but not simpler''. Economists
often adopt the first half of the sentence, ignoring the second half,
as an excuse for using oversimplified models, particularly for IAMs,
as climate models are often too complicated to be applied in IAMs
for economic analysis. Moreover, there is large uncertainty in future
temperature projections from climate models, as well as in future
economic systems. It is often hard to judge which model or scenario
is better, but policy makers have to make their decisions in the face
of the model and/or scenario uncertainty. In some simple cases, model
or scenario uncertainty can be represented as a special case of parameter
uncertainty. For example, \citet{GoulderMathai2000} compare two models
for carbon abatement knowledge accumulation: induced technical change
versus autonomous technical change. But these two energy models can
be connected with one single parameter: if the parameter's value is
0, then it represents the case of autonomous change, otherwise induced. 

There are many IAMs in the literature. They can be divided into two
broad categories: policy optimization IAMs and policy evaluation IAMs.
Policy optimization IAMs include a damage function mapping temperature
increases to economic damages, allowing the optimal policy to be found
using cost-benefit or cost-effectiveness analysis, so policy optimization
models are also called cost-benefit IAMs. Examples of policy optimization
IAMs include DICE \citep{Nordhaus_DICE2007,Nordhaus_DICE2016}, FUND
\citep{AnthoffTol2013}, PAGE \citep{hope_page09_2011}, WITCH \citep{bosetti_witch_2006},
MERGE \citep{Manne_merge_2005}, RICE \citep{Nordhaus_RICE_2010},
NICE \citep{dennig_inequality_2015}, DSICE \citep{CaiJuddLontzek2017_DSICE,CaiLontzek2019_DSICE},
and DIRESCU \citep{Cai_etal2019_DIRESCU}.\footnote{There are also many reduced form IAMs (e.g. \citealt{Golosov2014},
\citealt{jaakkola_non-cooperative_2019}, \citealt{brock_regional_2019}),
but they are mainly used for theoretical or qualitative analysis.
This review focuses on quantitative analysis for climate policies.}   These policy optimization IAMs are the main tools for calculating
the SCC. For example, DICE, FUND, and PAGE have been used by the US
Interagency Working Group to calculate the SCC \citep{IWG2010} under
different consumption discount rates.

Since policy optimization IAMs can only use a simple climate system
for computational tractability, some researchers developed policy
evaluation IAMs that assume that emissions or mitigation policies
are exogenous and have no feedback to the economy.\footnote{With advances in computational methods and hardware, IAMs are moving
forward by including more and more sectors as well as more complicated
climate system. For example, \citet{favero_forests_2020} apply the
dynamic global timber model \citep{sohngen_optimal_2003} with more
than 200 managed and natural forest ecosystems across 16 world regions
to assess the role of BECCS (Bioenergy with Carbon Capture and Storage)
in climate mitigation and also on economic outcomes such as timber
prices. \citet{Cai_etal2019_DIRESCU} study cooperative and noncooperative
climate policy with their dynamic stochastic IAM, DIRESCU, that incorporates
a more complicated climate system with heat and moisture transfer
between low latitude and high latitude regions.} Policy evaluation IAMs focus on quantifying future developmental
pathways and provide detailed information on the complex processes
in the carbon cycle, climate systems, land use, and/or other related
systems (sometimes economic systems), so they are also called process-based
IAMs or simulation IAMs. Examples of policy evaluation IAMs include
GCAM \citep{GCAM_2019}, IMAGE \citep{IMAGE_2014}, MESSAGE \citep{MESSAGE_2019},
AIM/CGE \citep{AIM_CGE_2017}, REMIND \citep{REMIND_2015}, and IGSM
\citep{chen_long-term_2016}.\footnote{In addition, agent-based models have also been employed in IAMs; see
\citet{farmer_third_2015} for a review.} These policy evaluation IAMs have been used to explore different
pathways for staying within climate policy targets, e.g., limiting
global mean temperature increase below 1.5 or 2 °C. See \citet{kelly1999integrated}
and \citet{weyant_contributions_2017} for more detailed discussion
about policy optimization IAMs and policy evaluation IAMs.

IAMs often assume some exogenous paths, such as population and technology
paths, but these paths are often uncertain, so these models have scenario
uncertainty. For example, policy evaluation IAMs often rely on exogenous
emission scenarios, but there are four representative concentration
pathways (RCPs) of greenhouse gas concentrations \citep{Meinshausen_RCP}:
RCP2.6, RCP4.5, RCP6, and RCP8.5, and they all have wide ranges in
2100. \citet{ONeill_etal2014} describe five Shared Socio-Economic
Pathways (SSPs) covering wide ranges of the projected population,
income, and temperature in 2100.

A different model or scenario provides a different optimal policy.
Usually policymakers do not know which model or scenario is more reliable,
and it is challenging to assign a belief distribution over the models
or scenarios. But they often have to choose only one policy. One way
is to do a multi-model or multi-scenario comparison and then find
some robust results from the models. For example, \citet{kim_assessing_2017}
analyze a set of simulations to assess the impact of climate change
on global forests under two emissions scenarios: a business-as-usual
reference scenario analogous to the RCP8.5 scenario, and a greenhouse
gas mitigation scenario, which is in between the RCP2.6 and RCP4.5
scenarios. \citet{Gillingham_etal_2018} compare six models: DICE,
FUND, GCAM, MERGE, IGSM, and WITCH. But in many cases, different models
or scenarios may lead to significantly different solutions, so it
is not possible to extract a robust policy from a multi-model or multi-scenario
comparison. 

A robust decision-making method, such as the max-min method or the
min-max regret method \citep{Iverson2012,Iverson2013,AnthoffTol2014,CaiSanstad2016},
is a tool to solve this problem in the face of model uncertainty.
For example, \citet{Cai_etal2017_FoodPolicy} apply the min-max regret
method to study robust decisions of agricultural research and development
in the face of uncertain SSP scenarios in population, income, and
temperature. \citet{rezai_climate_2017} derive max-min, max-max,
and min-max regret policies to deal with climate model uncertainty
(among DICE, FUND, and PAGE) and climate skepticism. If a distribution
for model uncertainty can be given, then model uncertainty aversion
can be incorporated. For example, \citet{Berger2017} study the impact
of model uncertainty aversion on optimal mitigation policy under catastrophic
climate risks.

\section*{Ambiguity and Misspecification}

If an uncertain parameter does not have a belief distribution, it
is ambiguous on its value, and approaches for dealing parameter uncertainty
can be applied to address the ambiguity. If there are many models
(or scenarios) but there is no belief distribution across the models
(or scenarios), the robust decision-making methods discussed in the
previous section can address the ambiguity on the choice of models
(or scenarios). An ambiguity-averse individual would rather choose
an alternative with a known probability distribution over one where
the probabilities are unknown. Thus, this review discusses only ambiguity
and misspecification on probability distributions.

A stochastic IAM often assumes that the probability distribution functional
form of a risk or shock is given with parameters estimated from historical
data, future projections, survey data, or expert opinions. For example,
researchers often assume that TFP is a lag-1 autoregression process
and its shock has a normal distribution with mean zero and an estimated
standard deviation. However, estimated parameters have standard errors,
implying that the true probability distribution is uncertain. Sometimes
even the functional form of the probability distribution may be misspecified.
For example, researchers provide many different belief distributions
for the climate sensitivity parameter \citep{IPCC2013}, but it is
unclear which particular one should be applied in IAMs. 

In the face of the probability ambiguity (i.e. deep uncertainty) and
misspecification, \citet{HansenSargent2008} introduce a robust control
framework with risk and ambiguity aversion, which is applied by \citet{Athanassoglou2012}
for an analytical pollution control problem. \citet{millner_scientific_2013}
study climate mitigation policy with ambiguity aversion and find that
the value of emissions abatement increases as ambiguity aversion increases.
\citet{Anderson_Robust_2018} conduct an empirically disciplined robustness
analysis for the size of the set of perturbations from their baseline
model of economic growth dynamics and climate dynamics. \citet{rudik_optimal_2019}
incorporates the robust control framework to include learning on climate
damage.\footnote{See \citet{brock_wrestling_2018} for a review on research challenges
in climate economics that focuses on three types of uncertainty: risk,
ambiguity, and misspecification.} \citet{baker_robust_2020} introduce a Robust Portfolio Decision
Analysis approach to help identify robust individual investments into
clean energy technology R\&D portfolios with deep uncertainty. \citet{barnett_brock_hansen_2020}
investigate risk, ambiguity, and misspecification with continuous-time
models and corresponding pricing methods to assess what sources of
uncertainty matter the most for the SCC. \citet{berger_are_2020}
find that policymakers are generally ambiguity averse.

\section*{Policy Uncertainty}

The preceding sections have focused on modeling and climate policy
analysis on pricing carbon in the face of uncertainty in the literature.
The models may provide quantitative estimates on the SCC, which is
often essential for regulatory policy evaluation and implementation,
but they often do not explicitly include climate policy instruments
to control emissions. In the literature, many climate policies have
been suggested, some of which have even been used globally, but there
is no consensus on what mix will be the best and how these policies
interact. This section discusses policy uncertainty in choice and
efficiency of climate policy instruments, innovation, geoengineering,
and adaptation. 

\subsection*{\textit{Climate Policy Instruments}}

There are many climate policy instruments, including carbon taxation,
cap-and-trade, intensity-based targets, and subsidies (for renewable
energy, research for new clean technology, emission reductions, etc.).
Each instrument has its advantages and disadvantages. For example,
a carbon tax gives a direct price on carbon emissions so companies
can adjust their emissions based on cost and benefit analysis, but
there is uncertainty in its effect on total emissions in the real
world. A cap-and-trade scheme issues a number of emission allowances
for the market to auction and trade, so it provides direct control
over future emissions and it would be more straightforward to control
temperature increase under some threshold (e.g. 2 or 1.5 °C), but
it is hard to estimate its economic cost. An intensity-based target
scheme requires emissions per unit of economic activity (e.g. output)
to not exceed given targets, so it may be appealing to developing
economies, but there is uncertainty in aggregate emissions and economic
costs. Subsidies to renewable energy can help renewable energy firms
to improve their market shares and competitivity with fossil fuel
energy firms, but there is uncertainty in its effect on controlling
total emissions. Carbon tax is the most popularly debated policy,
and it is often estimated to be equal to the SCC if it is not explicitly
modeled ( as in \citet{BaldwinCaiKuralbayeva}), and if emissions
control has not reached its limit \citep{CaiJuddLontzek2017_DSICE,CaiLontzek2019_DSICE}.
However, it could be challenging to politically pass a carbon tax
policy in some countries (such as the US). Instead, the cap-and-trade
scheme may be implemented at a regional level. For example, there
are currently cap-and-trade programs like the European Union Emissions
Trading Schedule, the Regional Greenhouse Gas Initiative, and the
California cap-and-trade program, though these programs require careful
design to make them effective.

Policy comparisons among the climate policy instruments have been
conducted in the literature. For example, \citet{goulder_instrument_2008}
review many instrument choices in climate policy with different evaluation
criteria, including economic efficiency and cost-effectiveness, distribution
of benefits or costs (across income groups, ethnic groups, regions,
generations, etc.), ability to address uncertainties, and political
feasibility. \citet{fischer_emissions_2011} compare carbon tax, cap
and trade, and intensity-based targets in a DSGE model with stochastic
productivity. \citet{heutel_2012} compares the optimal emissions
tax rate and the optimal emissions quota. \citet{Drouet2015} discuss
selection of climate policies under uncertainties. \citet{goulder_general_2016}
argue that under plausible conditions a more conventional form of
regulation, a clean energy standard, is more cost-effective than emissions
pricing such as carbon taxation or cap-and-trade. \citet{meckling_policy_2017}
investigate the combination and sequence of policies to avoid environmental,
economic, and political dead-ends in decarbonizing energy systems. \citet{rozenberg_instrument_2018}
compare the impact of mandates (for new power plants, buildings and
appliances), feebates (programs that tax energy-inefficient equipment
and subsidize energy-efficient equipment), energy efficiency standards,
and carbon pricing in a simple model with clean and polluting capital,
irreversible investment, and a climate constraint. They find that
carbon prices are efficient but can cause stranded assets, while feebates
and mandates do not create stranded assets. \citet{BaldwinCaiKuralbayeva}
compare a carbon tax with a subsidy for renewable energy using a DSGE
model, which is based on the full DICE model but adds renewable and
non-renewable energy sectors as well as a government who decides the
optimal dynamic carbon tax or subsidy. They find that a carbon tax
is more efficient under a stringent climate target, while a subsidy
is more efficient under a mild climate target.

In the recent literature, \citet{economides_monetary_2018} build
a New Keynesian DSGE model to explore how and to what extent monetary
policy should be adjusted under conditions of climate change. \citet{barrage_optimal_2020}
characterizes and quantifies optimal carbon taxes in a dynamic general
equilibrium climate\textendash economy model with distortionary fiscal
policy, and finds that optimal carbon tax schedules are 8\textendash 24\%
lower when there are distortionary taxes, compared to the setting
with lump-sum taxes considered in the literature. \citet{hafstead_designing_2020}
examine the role for tax adjustment mechanisms, which automatically
adjust the carbon tax rate based on the level of actual emissions
relative to a legislated target, and the trade-offs of alternative
designs. They show that tax adjustment mechanisms in carbon tax design
can substantially reduce emissions uncertainty. \citet{kalkuhl_all_2020}
find that the time-consistent policy is the ``all-or-nothing'' policy
with either a zero carbon tax or a prohibitive carbon tax that leads
to zero fossil investments, and it is the lobbying power of owners
of fixed factors (land and fossil resources), rather than fiscal revenue
considerations or the lobbying power of renewable or fossil energy
firms, that determines which of the two outcomes (all or nothing)
is chosen. \citet{van_der_ploeg_risk_2020} allow for immediate or
delayed carbon taxes and renewable subsidies that will cause discrete
jumps in the present valuation of physical and natural capital, and
then investigate how the legislative \textquotedblleft risk\textquotedblright{}
of tipping into policy action affects when the fossil era ends, the
profitability of existing capital, and green paradox effects \citep{sinn_public_2008}.

\subsection*{\textit{Innovation }}

The adoption of a regulatory policy to limit emissions may be the
necessary first step toward creating a market for low-carbon technologies.
But it should also be complemented by policies that specifically address
the technology innovation and development pipeline. Environmentally
friendly technologies to limit emissions are divided into three categories:
fossil fuel augmenting, alternative energy augmenting, and offset
technologies (i.e. carbon geoengineering). Fossil fuel augmenting
technologies improve the efficiency of fossil fuel use by allowing
more output per unit of fuel (e.g., hybrid or electric cars substitute
petroleum for electricity). Alternative energy augmenting technologies
such as solar panels or wind power generation utilize alternative,
non-emitting energy sources. Offset technologies directly reduce carbon
pollution, either at the point of emission via carbon capture by electricity
generators or by a distinct process that sequesters carbon from the
atmosphere (e.g. afforestation). 

\citet{acemoglu_environment_2012} show that sustainable growth can
be achieved with temporary taxes/subsidies that redirect innovation
toward clean input, and use of an exhaustible dirty resource helps
the switch to clean innovation under laissez-faire. \citet{gans_innovation_2012}
shows that a tighter emissions cap will reduce fossil fuel usage and
that this will diminish incentives to improve fossil fuel efficiencies,
but more stringent climate change policy may not increase incentives
to adopt and develop technologies that augment alternative energy
sources, while it will likely increase innovation in abatement technologies.
\citet{acemoglu_transition_2016} develop a microeconomic model of
endogenous growth where clean and dirty technologies compete in production
and innovation, and then characterize the optimal policy path that
makes heavy use of research subsidies as well as carbon taxes. \citet{fried_climate_2018}
finds that a carbon tax induces large changes in innovation, and the
innovation response increases the effectiveness of the policy at reducing
emissions. \citet{gillingham_cost_2018} review the costs of various
technologies and actions aimed at reducing greenhouse gas emissions.
\citet{mccollum_energy_2018} find that low-carbon investments should
overtake fossil investments globally by around 2025 to meet the ``well
below 2\LyXThinSpace °C'' target of the Paris Agreement.

\subsection*{\textit{Geoengineering}}

Climate engineering is viewed as a way to control climate change.
It can be split into two broad categories: solar geoengineering and
carbon geoengineering. Solar geoengineering, also known as solar radiation
management, is a technology that reflects a small fraction of sunlight
back into space or increases the amount of solar radiation that escapes
back into space to cool the planet. Carbon geoengineering, often also
called carbon dioxide removal, is a technology to remove carbon dioxide
from the atmosphere. It includes afforestation, bio-energy with carbon
capture and sequestration, and direct carbon removal and storage.

\citet{tavoni_forestry_2007} show that forestry is a determinant
abatement option and could lead to significantly lower policy costs,
and avoided deforestation in tropical-forest-rich countries can crowd
out some of the traditional abatement in the energy sector and lessen
induced technological change in clean technologies. \citet{heutel_climate_2016,Heutel_etal2018}
and \citet{keith_solar_2017} investigate the use of solar geoengineering
as a substitute for emissions abatement to reduce the atmospheric
carbon burden. \citet{favero_using_2017} recommend using forests
to mitigate greenhouse gases by storing carbon and supplying woody
biomass for burning in power plants with carbon capture and storage.
\citet{griscom_natural_2017} estimate natural climate solutions including
conservation, restoration, and improved land management actions that
increase carbon storage and/or avoid greenhouse gas emissions across
global forests, wetlands, grasslands, and agricultural lands. \citet{proctor_estimating_2018}
estimate the global agricultural effects of solar radiation management
for managing global temperatures by scattering sunlight back to space.
\citet{abatayo_solar_2020} study the governance of solar geoengineering
using a laboratory experiment, and find that too much geoengineering
can occur, leading to considerable economic losses and increased inequality
between countries.

\subsection*{\textit{Adaptation}}

The preceding parts have discussed mitigation policies for controlling
climate change. Adaptation usually has little control on climate change,
so it is outside the scope of this review. Here present a brief review
of recent research work on adaptation, as it plays a critical role
in climate policy analysis. For instance, \citet{barreca_adapting_2016}
examine the evolution of the temperature-mortality relationship in
the US to identify potentially useful adaptations, and find that residential
air conditioning contributes a substantial fraction of the welfare
gains. \citet{burke_adaptation_2016} exploit large variation in recent
temperatures and precipitation trends to identify adaptation to climate
change in US agriculture, and find that longer-run adaptations appear
to have mitigated less than half\textemdash and more likely none\textemdash of
the large negative short-run impacts of extreme heat on productivity.
\citet{cai_climate_migration_2016} find a positive and statistically
significant relationship between temperature and international outmigration
only in the most agriculture dependent countries. \citet{chen_coastal_2018}
find that coastal soil salinity on crop production has direct effects
on internal and international migration in Bangladesh even after controlling
for income losses. \citet{gopalakrishnan_climate_2018} analyze coastal
climate change adaptation in the face of sea level rise, ocean warming
and acidification, and increased storminess, and conclude that adaptation
will require governance coordination across multiple levels. \citet{massetti_measuring_2018}
examine methods for measuring climate adaptation using the empirical
evidence. \citet{Cai_etal2019_DIRESCU} find that low-latitude regions
will implement stronger adaptation than high-latitude regions, and
adaptation can significantly reduce the optimal regional carbon tax.

\section*{}

\section*{Conclusion and Further Research}

I have reviewed state-of-the-art work on different types of uncertainty
in controlling climate change: parameter uncertainty, risk, model
uncertainty, scenario uncertainty, policy uncertainty, ambiguity,
and misspecification. Uncertainty often plays an essential role in
models and changes results significantly. With advanced computational
methods and hardware, it becomes possible to analyze policies in more
complex and realistic IAMs with uncertainty.

With advances in understanding the physical science of climate change
and the economic system, there are a number of potential future studies
that can incorporate uncertainty in climate change economics. \citet{burke_opportunities_2016}
discuss some research opportunities in climate change economics, particularly
in three areas: SCC refinement, policy evaluation, and evaluation
of climate impact and policy choices in developing countries. Incorporating
uncertainty into related research may generate interesting results.
Other research opportunities include richer and more realistic representations
of the economic and climate systems as well as policies with uncertainty:
spatial disaggregation as in \citet{KrusellSmith2017}, disaggregation
of intertemporal agents (overlapping generations) as in \citet{kotlikoff_making_2019},
disaggregation of sectors (e.g. adding the green finance sectors),
disaggregation of heterogeneous agents, integration with other systems
(e.g. the water system), and more realistic international trade and
international agreements in climate policies. It will be interesting
to incorporate uncertainty in research work on IAMs with climate impact
on income inequality, regional inequality, social conflict, human
health and ecosystems, and migration. As suggested by \citet{irwin_welfare_2016},
it will also be important for future work to evaluate sustainability
and resilience in the face of uncertainty and climate change.

\section*{Disclosure Statement}

The author is not aware of any affiliations, memberships, funding,
or financial holdings that might be perceived as affecting the objectivity
of this review.

\section*{Acknowledgements}

I acknowledge support from the National Science Foundation grants
SES-1463644 and SES-1739909, and the United States Department of Agriculture
NIFA-AFRI grant 2018-68002-2793. I would like to thank William Brock,
Kenneth Judd, Thomas Lontzek, Brent Sohngen, Anastasios Xepapadeas
and anonymous reviewers for their helpful comments. I thank the editorial
committee, especially Brent Sohngen, for the invitation to write this
review.

\section*{Further Reading}

Burke, Marshall, Solomon M. Hsiang, and Edward Miguel. 2015. \textquotedblleft Global
non-linear effect of temperature on economic production.\textquotedblright{}
\textit{Nature} 527 (7577):235\textendash 239.

\noindent Cai, Yongyang. 2019. \textquotedblleft Computational methods
in environmental and resource economics.\textquotedblright{} \textit{Annual
Review of Resource Economics} 11:59\textendash 82.

\noindent Cai, Yongyang and Thomas S. Lontzek. 2019. \textquotedblleft The
social cost of carbon with economic and climate risks.\textquotedblright{}
\textit{Journal of Political Economy} 6:2684\textendash 2734.

\noindent IPCC. 2013. \textit{Climate Change 2013, The Physical Science
Basis}. New York: Cambridge University Press.

\noindent Nordhaus, William D. 2008. \textit{A Question of Balance:
Weighing the Options on Global Warming Policies}. Yale University
Press.

\noindent Pindyck, Robert S. 2013. \textquotedblleft Climate Change
Policy: What Do the Models Tell Us?\textquotedblright{} \textit{Journal
of Economic Literature} 51 (3):860\textendash 872.

\noindent Stern, N. H. 2007. \textit{The economics of climate change:
the Stern Review}. Cambridge, UK: Cambridge University Press.

\noindent Weitzman, Martin L. 2011. \textquotedblleft Fat-Tailed Uncertainty
in the Economics of Catastrophic Climate Change.\textquotedblright{}
\textit{Review of Environmental Economics and Policy} 5 (2):275\textendash 292.

\bibliographystyle{jpe}
\bibliography{Cai_ORE_revision}

\end{document}